# Quantum Models of Mind: Are They Compatible with Environment Decoherence?

Luiz Pinguelli Rosa[1], Jean Faber[2]

[1]*Alberto Luiz Coimbra Institute for Graduate Studies and Research in Engineering, CT, COPPE, University of Rio de Janeiro, RJ, Brazil, lpr@adc.coppe.ufrj.br.*

[2]*National Laboratory of Scientific computing - LNCC, Quantum Computing Group, Petrópolis, RJ, Brazil, faber@lncc.br.*

**Abstract**

Quantum models of mind associate consciousness to coherent superposition of states in the brain. Some authors consider consciousness to be the result of a kind of internal quantum measurement process in the brain. In this paper we discuss the ideas of Hameroff - Penrose and Tegmark and their calculation for an estimate of decoherence time. We criticize the Hameroff - Penrose model in the context of quantum brain model by gravitational collapse orchestrated objective reduction (orch. OR), assumed by Penrose, and we propose instead that the decoherence process is caused by interaction with the environment. We consider it useful to exploit this possibility because of the growing importance of decoherence theory in quantum measurement, and also because quantum mechanics can be applied to brain study independently of the Hameroff -Penrose model for mind and consciousness. Our conclusion is that the Hameroff-Penrose model is not compatible with decoherence, but nevertheless quantum brain can still be considered if we replace gravitational collapse orch .OR with decoherence. However our result does not agree with Tegmark´s conclusion of refuting not only the Hameroff-Penrose gravitational collapse but also quantum brain, based on decoherence time calculations in specific cases in the brain. In spite of this fact we also disagree with some points of the response to Tegmark´s article given by Hagan, Hameroff and Tuszynski.

**1. Introduction**

Different works on the brain and mind problem have used quantum theory to explain the emergence of consciousness [1-8]. There are in quantum theory as well as in statistical physics collective phenomena, not reducible to individual components of the system. The conjecture is that collective quantum phenomena produce coherent states in the brain. As we shall see in this paper, decoherence was not taken into account generally in current quantum models of the mind until very recent polemical works [1,9].

Hameroff and Penrose [2,3] formulated what they call orchestrated objective reduction (orch. OR) in the brain, by extending the objective interpretation of quantum measurement as a gravitational physical mechanism, suggested by Penrose [5] instead of decoherence.

In this paper we discuss the Hameroff-Penrose model [2,3] and Tegmark´s arguments [9] against the orch. OR. We consider the possibility of substituting the idea of gravitational collapse (orch. OR) by the natural decoherence process to produce consciousness. We concentrate our discussion on quantum aspects of the brain. Hameroff and Penrose assume a quantum computing process in microtubules, which constitute the skeleton of neural cells. We show a mistake concerning the temperature regime of the decoherence process which was considered by Hameroff and Tuszynski [1] in their criticism of Tegmark´s calculation of decoherence rate. In section 2 we present a formal view of the constitution of the microtubules in the brain, as well as of the Hameroff -Penrose quantum model of consciousness production. In section 3 we point out some problems of this model and we discuss Tegmark´s critical paper [9]. Our calculation of decoherence time in the brain system, by assuming ion and dipole interactions, are in section 4. The conclusion and final comments are presented in section 5.

## 2. Microtubules and Quantum Process

Microtubules are hollow cylinders comprised of an exterior surface with a cross-section diameter of *25nm* with 13 arrays of protein dimers called tubulines. The interior of the cylinder contains ordered water molecules, which implies the existence of an electric dipole moment and an electric field [10-14]. The microtubules comprise the internal scaffolding (the cytoskeleton) in all cells including neurons. Cytoskeletal structures are determinant of both structure and function and are dynamically active, performing activities which are instrumental to cellular organization [10-14].

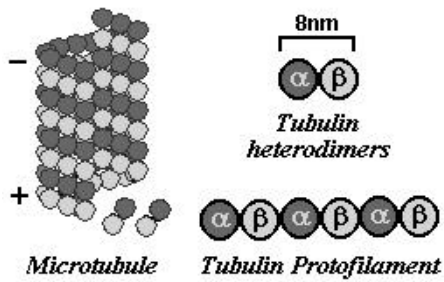

Fig 1 – Illustration of microtubule, tubulin and protofilament

The walls of microtubules in the cytoskeleton of neurons can work as cellular automata, able to store information [15] and to make computation by using combinations of the two possible states (as dimers) of the tubulins that constitute these walls [1,11-14]. The interior of the microtubule works as an electromagnetic wave guide, full of water in organized collective state, able to transmit information through the brain. A gelatinous state of water in brain cells, which was observed by Watterson [16], favors macroscopic collective effect of coherent quantum state at microscopic level. The conjugation of these possibilities can result in a coherent superposition of quantum states, embracing a very large number of microtubules.

Consciousness in the Hameroff-Penrose quantum model, using the microtubules structure, is produced in two steps. While the coherent superposition expands, there is a quantum computation through tubulins, such as dimers which can work as cellular automata in the walls of microtubules [15], and a propagation of information by wave guides inside microtubules of neurons. In this part of the cycle, which the coherent superposition of quantum states persists, there is a pre-consciousness state to the person. The second step is just the orchestrated reduction which produces consciousness [1-5].

## 3. Quantum Brain and Decoherence

A possibly vulnerable point in the quantum models of consciousness is the interaction between system and environment. According to the decoherence theory, macroscopic objects obey quantum mechanics. The interaction with the environment in this theory causes decoherence, which destroys quantum effects of macroscopic objects [17-19]. For this reason, these objects behave according to classical physics and Newton mechanics in an approximation of quantum theory. But for us this natural time limit for quantum superposition, generated by decoherence, is the fundamental process that yields consciousness. If we conciliate decoherence effects and the biological theory of Fröhlich [10,20], we can construct an interesting model to quantum mind-brain process. The Fröhlich theory describes quantum superposition for bio-macromolecules by considering the polarization and vibration properties of these molecules. Vibrations in polarized systems generate electromagnetic fields which can mediate interactions among the units of the system. The non-linearity of his model induces an energy transference among the vibration modes. This process can provide a condensation of the energy and consequently coherence in living systems [10,20].

The questions to be answered are: Can we deny the hypothesis of gravitational effect and maintain Hameroff and Penrose´s model to explain consciousness? In this hypothesis, can we use this model with decoherence? If we answer "yes" to the above questions we invalidate Penrose´s hypothesis of non-algorithmic theory which explains the emergent properties of the mind. In order to make a comparison with the predictions of Hameroff and Penrose, we discuss the decoherence time in the brain calculated by Tegmark [9], whose conclusion was to discard the relevance of the quantum approach to the brain. The period of time in which coherence is sustained in the brain was estimated at 500 ms by Hameroff and Penrose [2,3] based on biological time of brain response to external stimuli. This period is many times greater than the time of decoherence, considered an extremely efficient and fast natural process. The time of reduction by gravitational effect, calculated by Penrose, for water drops $10^{-4}$ cm and $10^{-3}$ cm in diameter is respectively 0.1s and $10^{-6}$ s. But typical decoherence time is estimated to be much shorter. They are $10^{-23}$ s and $10^{-9}$ s respectively for systems with a radius of $10^{-3}$ cm and $10^{-5}$ cm [18,22]. Hameroff and Penrose [3] have calculated the number of tubulins under coherent superposition assuming a coherence time *t* based on typical response time of the brain to external stimulus. The corresponding critical energy for orch. OR is related to time *t* by quantum uncertainty relation.

$$E = \frac{\hbar}{t}$$

(1)

The numerator is the Planck constant $(6.6260755 \times 10^{-34}$ Js$)$ over $2\pi$ and the denominator is the time $t$ which gives the order of magnitude. The critical energy $E$ is given by the gravitational self energy.

## 4. Comparisons of the Decoherence Rates

We assume, in place of gravitational collapse, that coherence grows as the system is isolated by some mechanism until coherent superposition reaches a critical situation, at which point decoherence is activated. When the limit given by the critical value of some physical parameter is reached, interaction with the environment becomes sufficient to cause decoherence and, therefore, to destroy the coherent collective state. The critical size has no connection to the space time structure. Therefore, the quantum system remains protected against decoherence by interaction with the environment, as isolated systems up to reach the critical size.

We can make a first rough estimate of t based on a macroscopic thermodynamic prescription of the critical energy, used in place of gravitational auto interaction in the Hameroff and Penrose [2,3] model. However, we can examine the problem from another point of view by making a microscopic quantum description. Tegmark [9] has taken into account, in a detailed model, the relevant degrees of freedom in the brain. His conclusion is that the system is not sufficiently shielded from environmental influence to maintain quantum coherence in such a way as to allow quantum computation. He considers as limits two scales: a macroscopic one for superposition of neurons firing and a microscopic one for microtubules. In the first one he estimates $t = 10^{-20}$ s and in the second one $t = 10^{-13}$ s. To calculate the latter value Tegmark considers the propagation along the microtubule, in the direction of its axis, $z$, of a kink-like perturbation which changes the tubulin electric dipole from $+p_0$ to $-p_0$:

(2) $$p(z) = \begin{cases} +p_0 \ldots z \gg z_0 \\ -p_0 \ldots z \ll z_0 \end{cases}$$

For this calculation the author uses the total charge generated by positive ions of calcium in a ring of tubulins at the cross-section of the microtubule, at the point of the kink-like propagation. This charge, distributed around the microtubule wall at the point of the perturbation, interacts with distant ions of the environment through Coulomb potential. The distributed charge is treated as one particle and the environmental ion acts as the other particle. Several approximations are made. The Coulomb potential $R^{-1}$ is expanded in a Taylor series around R up to the second order term. The decoherence effect is produced by the term of the Taylor expansion with second order derivative of the potential, which is proportional to $R^{-3}$, as one observes in tidal force exerted by the moon´s gravitational attraction.

The formula used by Tegmark for the decoherence time of microtubule quantum states generated by interaction with environmental ions is:

(3) $$t = \frac{R^3 \sqrt{M \kappa T}}{K N e^2 s}$$

where K is the Coulomb constant, R is the distance between the interacting microscopic system in the tubulin and the environmental ion, M is the mass of the ion, N is the number of elementary charges in the microtubule interacting system, s is the maximal "separation" between the position of two kinks in quantum superposition, $e$ is the electron charge, $\kappa$ is the Boltzmann constant and T is the temperature. If we consider the values of the physical constants K = $9 \times 10^9$ N m$^2$ C$^{-2}$, $\kappa = 1.38 \times 10^{-23}$ J $^0$K$^{-1}$ and $e = 1.6 \times 10^{-19}$ C and taking T = 309 $^0$K, M the mass of a water ion, 18x *proton mass* $(1.67 \times 10^{-27}$ kg$)$, N = $10^3$ estimated by Tegmark, who assumed R and s with the order of magnitude of the microtubule diameter $24 \times 10^{-9}$ m, we have from formula (3) the order of magnitude $t = 10^{-13}$ s.

| *Quantum superposition of* | *Decoherence time calculed* |
|---|---|
| Superposition of neural firing [4] | $10^{-20}$s |
| Soliton superposition [4] | $10^{-13}$s |
| Orch. OR superpositions [1] | $10^{-5}$s - $10^{-4}$s |
| Decoherence Model (MT - ion interaction) | $10^{-9}$s |
| Decoherence Model (MT - dipole interaction) | $10^{-16}$s |

Table 1 – Calculed values of the decoherence time scale

Hagan, Hameroff and Tuszynski [1] in their response to Tegmark´s paper agreed with his result at the level of neuron, in the macroscopic scale, but not in the microscopic scale where tubulins and microtubules play a role. One reason is that Tegmark´s model takes kink-like soliton waves [23] along the microtubule, while orch. OR considers superposition of different conformational states of a tubulin dimer [1]. In spite of their criticism, Hagan, Hameroff and Tuszynski have used Tegmark model with some modifications, using electric dipole momentum in the place of Coulomb potential to describe tubulin interaction with environmental ion:

(4) $$\hat{o} = \frac{R^4 \sqrt{I} \, \hat{e}\hat{O}}{3\hat{E}eps} \hat{U}_{dipole}$$

where R is the distance between the environmental ion and the dipole generated by the tubulin of the microtubule, p is the electric dipole momentum, the other variables are analogous to those of formula (3), unless the last factor in (4), which is a geometric factor involving the angles between the directions of the line from the dipole to the ion, of the dipole and of the separation. It is assumed to be of order of 1 [1]. In their approximation the authors claim that the decoherence time lies in the range $10^{-6}$ s - $10^{-4}$ s.

However if we apply formula (4) with R, M and s having the same order of magnitude of the values used by Tegmark, but using for the tubulin dipole momentum [1] considering at the direction of microtubule axis the value $p = 10^{-27}$ Cm, we go back to our previous result $t = 10^{-10}$ s.

## 5. Results

The authors have also criticized formula (3) used by Tegmark [9] because it predicts that the decoherence time increases with the square root of the temperature. We can see from very general formulas (1) and (2) that time $t$ decreases when temperature T increases. On the contrary, according to formula (3) of Tegmark at low temperature, where interaction with the environment must be minimal, the decoherence is faster, in contradiction with experience. Coherent macroscopic states, such as superconductivity, happen at temperature approaching zero. We show in what follows that Tegmark´s formula is not valid at low temperature. This is not the case of the human body but it shows that Tegmark's estimates cannot be used generally. That is, the formulas (3) and (4) are just an approximation for a right regime of temperature.

To discuss Tegmark formula (3), let us consider that the density matrix for the positions of the two particles 1 (charged macromolecule) and 2 (environmental ion) interacting through a potential V(r). From Schrödinger equation for density matrix $r$ the system evolves as:

(5) $$\frac{d\mathbf{r}}{dt}(r,t) = [H, \mathbf{r}(r,t)]$$

where H is the Hamiltonian of system. In this case we assume H = V(r), therefore we have:

(6) $$\mathbf{r}(\vec{r}_1, \vec{r}_2, \vec{r}_1', \vec{r}_2', t) = \\ = \mathbf{r}(0) \exp\left[-\frac{i}{\hbar}(V(\vec{R}) - V(\vec{R}'))t\right]$$

where

(7) $$\begin{cases} \vec{R} = \vec{r}_2 - \vec{r}_1 \\ \vec{R}' = \vec{r}_2' - \vec{r}_1' \end{cases}$$

Decoherence drops the non diagonal part of macro particle 1 reduced density operator due to the interaction with particle 2, when we assume that the system 1 and 2 start in a tensorial product state $\mathbf{r} = \mathbf{r}_1\mathbf{r}_2$, the density operator is averaged by tracing over the environment. The result is:

(8) $$\mathbf{r}_1(\vec{r}_1, \vec{r}_2, t) = Tr(\mathbf{r}(\vec{r}_1, \vec{r}_2, \vec{r}_1', \vec{r}_2', t)) = \\ = \mathbf{r}_1(\vec{r}_1, \vec{r}_2, t) \exp(-\Lambda t) \quad for \quad \vec{r}_1 \neq \vec{r}_1'$$

That is, the non diagonal part vanishes for $t > \Lambda^{-1}$. To obtain this result, we choose:

(9) $$\begin{cases} \vec{r}_1 = (0,0,0) \\ \vec{r}_1' = (x_1, y_1, z_1) \\ \vec{r}_2 = \vec{r}_2' = (x_2, d, 0) \end{cases}$$

The first two relations mean a coordinate axis choice, while the last one means an approximation where the environmental ion (particle 2) moves along the x axis (d is a constant) and the change of direction of its motion is neglected in the interaction, assumed as too small. The shortest distance from it to the particle 1 is *d*. So:

(10) $$\begin{cases} \vec{R} = \vec{r}_2 \\ \vec{R}' = \vec{r}_2 - \vec{r}_1' \end{cases}$$

We can expand the Coulomb potential up to the second order term:

(11) $$V(\vec{R}') = Kq_1q_2 \frac{1}{|\vec{r}_2 - \vec{r}_1'|} \\ \cong Kq_1q_2 \left(\frac{1}{|\vec{r}_2|} + \frac{x_1 x_2 + y_1 d}{|\vec{r}_2|^3}\right)$$

and hence:

$$(12) \quad V(\vec{R}') - V(\vec{R}) = K q_1 q_2 \left( \frac{x_1 x_2 + y_1 d}{|\vec{r}_2|^3} \right)$$

where $q_1$ and $q_2$ are the charges of particles 1 and 2.

The figure below shows the relation MT-environment expressed in (10):

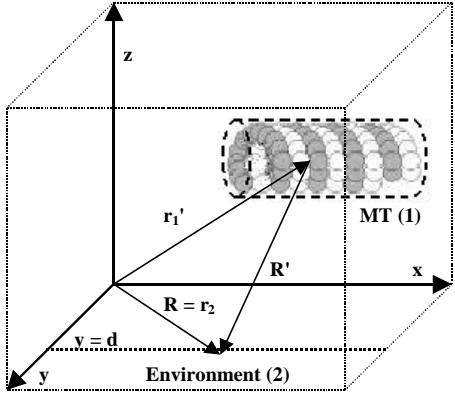

Fig 2 – Schematic illustration of the co-ordinates of the enviroment-MT system

The next step is to assume a separable form for the density matrix and a Gaussian distribution with zero mean and variance $\lambda\sqrt{2}$ for particle 2:

$$(13) \quad \rho_1(\vec{r}_1, \vec{r}_2, t) = \frac{\rho_1(0)}{\sqrt{4\pi}\lambda} \int \exp\left[-\left(\frac{x_2}{2\lambda}\right)^2\right] \times$$
$$\times \exp\left[-\frac{K q_1 q_2 t}{i\hbar}\left(\frac{x_1 x_2 + y_1 d}{(x_2^2 + d^2)^{3/2}}\right)\right] dx_2$$

For $d \gg x_2$ we can approximate $(x_2^2 + d_2^2)^{3/2} \cong d^3$ and with a appropriate algebraic manipulation the equation (13) becomes:

$$(14) \quad \rho_1 = C \exp\left[\frac{t y_1 K q_1 q_2}{i\hbar d^2} - \left(\frac{t x_1 \lambda K q_1 q_2}{\hbar d^3}\right)^2\right] \Phi$$

where

$$\Phi = \int_{-\infty}^{+\infty} \exp\left[-\left(\frac{x_2}{2\lambda} + \frac{i t x_1 \lambda K q_1 q_2}{\hbar d^3}\right)^2\right] dx_2; \quad C = \frac{\rho_1(0)}{\sqrt{4\pi}\lambda}$$

The last integration in (14) yields a constant number; the first imaginary term in the exponential is a phase factor, while the second one produces a decrease of the non-diagonal part of the density matrix with a characteristic decoherence time:

$$(15) \quad \tau = \frac{\hbar d^3}{x_1 \lambda K q_1 q_2}$$

if we assume thermal constraint $(\Delta p)^2/M \approx \kappa T$ and take uncertainty principle $\Delta x \approx \hbar/(M \kappa T)^{1/2}$ as Tegmark did [4], we rescue formula (3).

The square root of $\kappa T$ in the numerator of (15) depends on the approximation used $\lambda \leq d$ and $x_2 \ll d$. However, if we use a much broad function to represent the environmental ion in the density matrix, such that it vanish effectively for $x_2 = \lambda \gg d$ the situation is changed in the expression (13):

$$(16) \quad \rho_1(\vec{r}_1, \vec{r}_2, t) = C \int \exp\left[-\frac{1}{4} - \frac{K q_1 q_2 t}{i\hbar} \frac{x_1 \lambda + y_1 d}{(\lambda^2 + d^2)^{3/2}}\right] dx_2$$

The integral above will vanish if the imaginary exponential oscillates upwards too close to the limit, producing a cancellation between positive and negative parts of the function in the integration, which vanishes when:

$$(17) \quad t \gg \frac{\hbar}{K q_1 q_2} \frac{(\lambda^2 + d^2)^{3/2}}{x_1 \lambda + y_1 d} \cong \frac{\hbar}{K q_1 q_2} \frac{\lambda^2}{x_1}$$

By simple inspection we see that, with $\lambda = \hbar/(M \kappa T)^{1/2}$, the factor $\kappa T$ goes down to the denominator in (17), as usual, differently from the Tegmark formula (3):

$$(18) \quad \tau = \frac{1}{K q_1 q_2} \frac{\hbar^3}{x_1 M k T}$$

At very low temperature $\lambda$ becomes too high and the approximation used in formula (3) does not work. In this case we must use formula (18) instead. So we remove the obstacle put by Hagan, Hameroff and Tuszynski [1] against the validity of Tegmark formula.

For electric dipole case (4) we can use the same process used previously. So, let us consider a composite system of 2 particles, where the particle 1 is the electric dipole and the particle 2 represents the environment. Both of them interacting through a dipole potential V(r) which expanded:

$$V(\vec{R}') = Kq \frac{\vec{p} \cdot (\vec{r}_2 - \vec{r}_1')}{|\vec{r}_2 - \vec{r}_1'|^3}$$

(19)

$$\cong Kq \left( \frac{\vec{p} \cdot \vec{r}_2}{|\vec{r}_2|^3} + \frac{3\vec{p} \cdot \vec{r}_2 |\vec{r}_1'|}{|\vec{r}_2|^4} \right)$$

From (5) and taking all previous considerations, the density matrix of system becomes:

(20)

$$\rho_1(\vec{r}_1, \vec{r}_2, t) = C \int \exp\left[ -\left( \frac{x_2}{2\lambda} \right)^2 - \frac{3|\vec{r}_1'|Kq}{i\hbar} \left( \frac{p_x x_2 + p_y d}{(x_2^2 + d^2)^2} \right) t \right] dx_2$$

For a regime where d>>$x_2$ we can approximate $(x_2^2 + d_2^2)^2 \cong d^4$ and therefore:

(21)

$$\hat{\rho} = \frac{d^4 \sqrt{M\kappa T}}{3Kqps} \hat{U}_{dipole}$$

where we assume $p_x$= pcos($\alpha$) with $\Omega_{dipole}$ = sec($\alpha$) and |r| = s. Now, if we consider another regime with a broad function to represent the environmental, so that it vanishes effectively for $x_2$ = $\lambda$>>d, where $\lambda$ = ℏ/(M κT)$^{1/2}$:

(22)

$$\hat{\rho} = \frac{\hbar^4 (M\kappa T)^{-3/2}}{3Kqps} \hat{U}_{dipole}$$

Formula (22) shows that in the very low temperature regime our calculations for the dipole case yields a result compatible with the high decoherence time in the limit of very low temperature, as does Tegmark´s [9]. The result we obtained is summarized in table 1, where we compare all the values obtained from our estimate as well as other estimates.

Both are lower than the Tegmark [9] case and higher than Hagan, Hameroff and Tuszynski [1] for dipole approximation.

## 6. Conclusions

The result stated above points out that there is a big difference between the time period of coherence in the Hameroff-Penrose model [2,3], and the time of decoherence in the brain. We point out that their model must be modified when including decoherence in the place of gravitational collapse. However, based on this difference, we do not conclude, as Tegmark [9] does, that quantum approach to the brain problem is refuted if we use decoherence instead of gravitational collapse. A first point is that we must also consider the time for building coherence, while the system either remains relatively isolated to sustain coherence, otherwise there is no coherent collective state. Hameroff and Penrose speak about isolating microtubules from environmental entanglement but they did not take this point further.

We believe there is some kind of Fröhlich process acting in the brain sustaining the coherence [20]. In the Fröhlich theory the energy supplied to a nonlinear vibration system balanced by energy losses due to interaction with its surroundings is not immediately thermalized and can disturb thermodynamic equilibrium through energy condensation [10]. Depending on energy supplied and on energy transferred to the environment, the system shifts the regime from incoherent to time limited coherent and in further steps to steady coherent.

Our result does not discard the conjecture that quantum theory can help us to understand the functioning of the brain, and maybe also to understand consciousness. We assume that the neuronal network in the brain processes information coming from the external environment or stored in the memory, working at sub-neuronal level as a quantum system. Quantum coherent superposition of states is sustainable at the macroscopic level for a very short time, like a flash. When the physical system reaches a critical situation coherence is destroyed due to decoherence process by interaction with the environment.

In spite of our disagreement with Tegmark [9] concerning his refutation of quantum brain together with the Hameroff and Penrose orch. OR model [2,3], we have shown that our calculation does not agree with the response to Tegmark´s paper by Hagan, Hameroff and Tuszynski [1].

We still propose a new quantum model in the brain where the most important thing is the sequence of coherent states accumulating in the microtubule. In this manner, the quantum activity could appear in another formulation for the brain. The details of these considerations will be present in our future articles.